\documentclass[aps,prl,nofootinbib,twocolumn,superscriptaddress,preprintnumbers]{revtex4-1}
\usepackage[nodisplayskipstretch]{setspace}
\usepackage{cancel}
\usepackage{mathtools}
\usepackage{slashed}

\usepackage{amssymb}
\usepackage{amsmath}
\usepackage{epsfig}
\usepackage{hyperref}
\usepackage{breakurl}
\usepackage{bm}
\usepackage{color}
\usepackage{tikz}
\usetikzlibrary{decorations.markings}
\usepackage[english]{babel}
\DeclareMathOperator{\arccosh}{arccosh}

\newcommand\beq{\begin{equation}}
\newcommand\eeq{\end{equation}}
\def\bea{\begin{eqnarray}}
\def\eea{\end{eqnarray}}

\DeclareRobustCommand{\SkipTocEntry}[4]{}

\newcommand{\nn}{\nonumber}

\newcommand\beal{\begin{aligned}}
\newcommand\eeal{\end{aligned}}

\usepackage{xcolor}

\newcommand{\bp}{{\boldsymbol p}}

\newcommand{\br}{{\boldsymbol r}}


\begin{document}


\title{Dynamics of Binary Systems to Fourth Post-Minkowskian Order\\ [0.2cm] from the Effective Field Theory Approach}
\author{Christoph Dlapa}
\affiliation{ Deutsches Elektronen-Synchrotron DESY, Notkestr. 85, 22607 Hamburg, Germany}
\affiliation{Max-Planck-Institut f\"ur Physik, Werner-Heisenberg-Institute, 80805 Munich, Germany}
\author{Gregor K\"alin}
\author{Zhengwen Liu}
\author{Rafael A. Porto}
\affiliation{ Deutsches Elektronen-Synchrotron DESY, Notkestr. 85, 22607 Hamburg, Germany}

\begin{abstract}
  We present the contribution from potential interactions to the dynamics of non-spinning binaries to fourth Post-Minkowskian (4PM) order. This is achieved by computing the scattering angle to ${\cal O}(G^4)$ using the effective field theory approach and deriving the bound radial action through analytic continuation. We reconstruct the Hamiltonian and center-of-mass momentum in an isotropic~gauge. The  (three-loop) integrals involved in our calculation are computed via differential equations, including a sector yielding elliptic integrals. Using the universal link between potential and tail terms, we also report: {\it i)}~The instantaneous energy flux at~${\cal O}(G^3)$; {\it ii)}~The contribution to the 4PM unbound/bound radial action(s) depending on logarithms of the binding energy; {\it iii)} The (scheme-independent) logarithmic contribution to the 4PM non-local tail Hamiltonian for circular orbits. Our results in the potential region are in agreement with the recent derivation from scattering amplitudes. We also find perfect agreement in the overlap with the state-of-the-art in Post-Newtonian~theory.  
  \end{abstract}

\maketitle

\emph{Introduction.} Motivated by the success of current gravitational wave (GW) detectors and the planned empirical reach \cite{ET,LISA} and scientific output  \cite{buosathya,tune,music,Maggiore:2019uih,Barausse:2020rsu} of future GW observatories, there has been a significant amount of progress in applying ideas from particle physics --- such as the effective field theory (EFT) approach in~the Post-Newtonian (PN) \cite{nrgr,nrgrs,dis1,dis2,prl,nrgrss,nrgrs2,andirad,andirad2,amps,srad,chadRR,natalia1,natalia2,tail,apparent,lamb,nrgr4pn1,nrgr4pn2,tail3,5pn1,5pn2,hered,Blumlein:2020pyo,foffa3,blum,blum2,dis3,Levi:2020uwu,Levi:2020kvb,withchad,radnrgr,pardo,Cho:2021mqw,Cho:2022syn,Blumlein:2021txe,hered2,walterLH,foffa,iragrg,review} and Post-Minkowskian (PM) regimes \cite{pmeft,3pmeft,tidaleft,pmefts,janmogul,janmogul2,Mougiakakos:2021ckm}, as well as other tools from quantum field theory and scattering amplitudes, e.g.~\cite{ira1,Vaidya:2014kza,Walter,Goldberger:2016iau,cheung,Guevara:2018wpp,donal,donalvines,cristof1,bohr,zvi1,zvi2,paper1,paper2,Haddad:2020que, Aoude:2020onz,parra,zvispin,soloncheung,andres2,4pmzvi,Parra2,Gabriele,DiVecchia:2021bdo} --- to the study of the inspiral problem with compact objects in general relativity. Most of these developments have impacted our knowledge of the conservative dynamics of the two-body problem in gravity, reaching the next-to-next-to-next-to-next-to leading order (N$^4$LO) \cite{tail,apparent,lamb,nrgr4pn1,nrgr4pn2} for spin-independent contributions in the PN regime, complementing traditional methodologies \cite{blanchet,Damour:2014jta,Jaranowski:2015lha,Bernard:2015njp, Bernard:2017bvn,Marchand:2017pir}. The~hunt for accuracy continues, with the state-of-the-art in PN theory reaching towards the N$^5$LO at 5PN \cite{5pn1,5pn2,Blumlein:2020pyo,hered,Blumlein:2021txe,hered2}, with partial results also at 6PN order \cite{blum,blum2,binidam1,binidam2}, in addition to the well-established breakthrough result at N$^2$LO (3PM) \cite{zvi1,zvi2,3pmeft} in the PM counter-part.\vskip 4pt

In this letter, we present the contribution from potential modes to the dynamics of non-spinning binary systems at ${\cal O}(G^4)$ using the EFT formalism \cite{pmeft,3pmeft}, in conjunction with the  Boundary-to-Bound (B2B) dictionary \cite{paper1,paper2,b2b3}. Our derivation proceeds through the scattering angle, which we compute via Feynman diagrams. Using several powerful multi-loop integration tools from particle physics \cite{Chetyrkin:1981qh, Tkachov:1981wb,Cheng:1987ga,Kotikov:1991pm,Remiddi:1997ny,Lee:2012cn,Smirnov:2012gma,Henn:2013pwa,Lee:2014ioa,Meyer:2016zeb,Meyer:2016slj,Prausa:2017ltv,Broedel:2019kmn,Primo:2017ipr,Smirnov:2019qkx,Smirnov:2020quc,Lee:2020zfb},  the calculation is reduced to a series of `three-loop'  integrals. The latter are obtained to all orders in velocities through differential equations, yielding {\it multiple polylogarithms} (MPLs) \cite{Chen:1977oja,Goncharov:2001iea,Duhr:2014woa,Duhr:2019tlz}, as well as complete elliptic integrals. The resulting deflection angle agrees with the potential-only result in~\cite{4pmzvi}, confirming its validity with an independent calculation entirely within the classical domain.\vskip 4pt

Armed with the solution of the scattering problem, observables for bound orbits follow from the B2B correspondence, after analytic continuation in the binding energy and angular momentum \cite{paper1,paper2}. We also reconstruct the Hamiltonian and center-of-mass momentum in an isotropic gauge. However, because of the separation into regions, we encounter spurious infrared(IR)/ultraviolet(UV) divergences \cite{tail,apparent}, similar to what occurs in the Lamb~shift~\cite{lamb}. As expected, these IR/UV poles ultimately disappear from observable quantities \cite{nrgr4pn2}. Yet, it makes the result from potential modes dependent on the regularization scheme, featuring for instance logarithms of the renormalization scale in dimensional regularization, as well as scheme-dependent finite pieces, which must cancel out in the final answer \cite{apparent,nrgr4pn2}.\vskip 4pt  The split into regions, however, turns out to be useful for several reasons. First of all, after removing some spurious contributions, the potential-only result serves as a proxy to the full answer. In addition, the universal link between potential and tail terms \cite{tail3,b2b3}, together with the nature of the integrals involving (on-shell) radiation modes, e.g.~\cite{ DiVecchia:2021bdo}, allows us to immediately read off several new results involving radiative effects. In particular, we report the leading logarithms (depending on the binding energy) in the unbound/bound radial action(s) to all orders in velocities. Moreover, we also present the instantaneous flux at~${\cal O}(G^3)$. In combination with the conservative part, the latter allows us to derive the GW phase entirely within a PM scheme for the first time. From the flux we can also read off the logarithmic contribution to the (non-local) tail Hamiltonian, which we compute for circular orbits.\vskip 4pt  We~have checked that the potential-only result as well as those involving the universal (non-local-in-time) radiation-reaction effects are consistent with the state-of-the-art in PN theory \cite{Blumlein:2020pyo,binidam2}.  The complete derivation of tail terms is presented in \cite{tail4pm} and \cite{zvitail} using EFT- and amplitude-based methodologies, respectively.\vskip 4pt

{\it The EFT formalism.}  The effective action, $S_{\rm eff}[x_a^\alpha(\tau_a)]$ $(a=1,2)$,  is obtained in the weak-field regime, $g_{\mu\nu} = \eta_{\mu\nu}+ h_{\mu\nu}$,  by \emph{integrating out} the metric perturbation in the (classical) saddle-point approximation via Feynman diagrams. The compact bodies are described by a worldline theory, which may be taken linear in the metric for non-spinning structureless bodies~\cite{pmeft}. As usual, integration in $d=4-2\epsilon$ dimensions is used to handle divergent integrals. The equations of motion (EoM) follow by extremizing $S_{\rm eff}[x_a^\alpha(\tau_a)]$. The scattering problem is then tackled by solving for the total impulse, $\Delta p_a^\mu$, and subsequently the deflection angle, \beq
2\sin(\chi/2) = \sqrt{-\Delta p_a^2}/p_\infty\,,\label{angle}
\eeq 
with $p_\infty$ the incoming center-of-mass momentum. Due to the presence of intermediate divergences we will use the PM expansion, with $\bar\mu^2 \equiv 4\pi \mu^2 e^{\gamma_E}$ in dimensional regularization,
\beq
\frac{\chi}{2} =  \sum_n \left(\left(4\bar\mu^2b^2\right)^\epsilon \frac{GM}{b}\right)^n \chi^{(n)}_{b }(\gamma)\,.\label{pmexp}
\eeq
We restrict to the impulse in the direction of the impact parameter, $b=\sqrt{-b^\mu b_\mu}$, which is sufficient to reconstruct the  answer using momentum conservation \cite{pmeft}. \vskip 4pt

In principle, we may tackle the entire (classical) {\it soft} region at once, without splitting into potential and radiation modes. However, even though it introduces spurious divergences \cite{apparent}, mode factorization remains a convenient tool to compute the integrals and isolate various contributions. In fact, the cancelation of intermediate poles not only becomes a non-trivial consistency check, it also allows us to extract several new results including radiation modes, using the universal character of the answer~\cite{b2b3}. We discuss radiation-reaction conservative effects in more detail in \cite{tail4pm}. \vskip 4pt
 
\begin{figure}[t!] 
\includegraphics[width=0.35\textwidth]{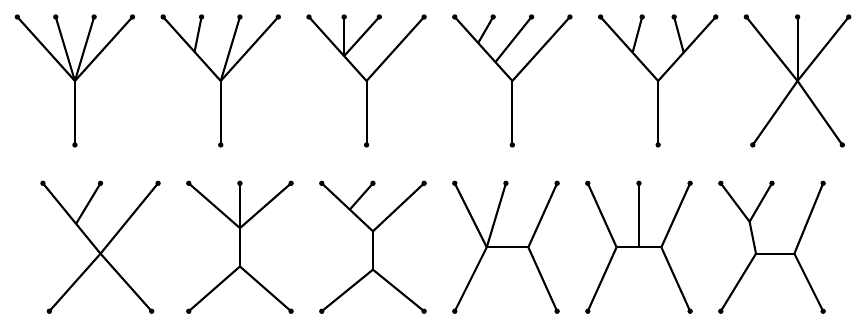} 
      \caption{Feynman topologies in the potential region at 4PM.} 
      \label{pot}
            \vspace{-0.4cm}
\end{figure}

{\it Potential region.}  The Feynman topologies needed at 4PM are shown in Fig.\,\ref{pot}. We must also include {\it iterations} of the deflected trajectories into lower order contributions to the effective action. The additional Feynman topologies are shown in \cite{3pmeft}. After tensor decomposition, and prior to Fourier transforming from transfer-momentum,~$q$, onto impact-parameter space \cite{pmeft, 3pmeft}, the impulse becomes a linear combination of scalar (three-loop) relativistic integrals of the~type
\begin{align}\label{4pm-ints}
\prod_{i=1}^3 \int \frac{\mathrm{d}^d \ell_i}{\pi^{d/2}}
\frac{\delta(\ell_i \!\cdot\! u_{a_i})}  { 
(\pm \ell_i \!\cdot\! u_{\not{a}_i} {-} i0^+)^{n_i}} {1 \over \prod_{j=1}^{9} D_j^{\nu_j}}\,,
\end{align}
with $n_i,\nu_j \in \mathbb{Z}$, $a_i\in\{1,2\}$ ($\slashed{1}=2,\slashed{2}=1$), $u_a$ the incoming velocities, and $D_j$ various sets of square propagators. The external data~obeys $q\cdot u_1 = q\cdot u_2 = 0$, $u_a^2=1$, such that the result depends on $t=-q^2$ and $\gamma\equiv u_1\cdot u_2$. As~before \cite{3pmeft}, the $t$ dependence follows from dimensional analysis. Remarkably, all our integrals can be classified into two families of square propagators. 

A~salient aspect of the EFT integrals in \eqref{4pm-ints} is that they are accompanied by delta functions, $\delta(\ell_i \cdot u_a)$, arising after time integration over the worldline sources. This feature, which is directly connected to the mass scalings in the total impulse \cite{paper1,pmeft,3pmeft}, yields a natural classification into sectors. We~denote as $(abc)$ the product $\delta(\ell_1 \cdot u_a) \delta(\ell_2 \cdot u_b) \delta(\ell_3 \cdot u_c)$. Integrals in sector $(111)$ (or its $(222)$ mirror), arising from the first five diagrams in Fig.\,\ref{pot} and lower-order iterations, capture the test-particle limit. These can be readily obtained by resolving the delta functions in the rest-frame, leading to three-dimensional {\it static} integrals. 
The static sector may be further decomposed in a basis of seven integrals via integration by parts (IBP) identities \cite{Chetyrkin:1981qh, Tkachov:1981wb},  using \texttt{FIRE6} \cite{Smirnov:2019qkx,Smirnov:2020quc} and \texttt{LiteRed} \cite{Lee:2012cn}. The associated master integrals are computed via standard tools, such as parametric representations and Cheng-Wu theorem~\cite{Cheng:1987ga}. We have also checked them numerically using \texttt{pySecDec}~\cite{Borowka:2017idc}. \vskip 4pt

The remaining integrals involving the last seven diagrams in Fig.\,\ref{pot}, as well as non-static iterations, can be reduced to solving the $(112)$ sector, for which we implemented the method of differential equations \cite{Kotikov:1991pm,Remiddi:1997ny}. Following \cite{parra,3pmeft}, we derived a system of equations with respect to the parameter $x$, defined via the relation $\gamma \equiv (x^2 {+} 1)/2x$. When possible, we solved the differential equations by means of a canonical basis \cite{Henn:2013pwa}. We use the algorithms described in \cite{Lee:2014ioa, Prausa:2017ltv, Lee:2020zfb}, yielding 
\begin{align}\label{DE-canonical}
\mathrm{d}\vec{f}(x,\epsilon) =\epsilon \Big(\sum_i M_i\, \mathrm{d}\log \alpha_i(x) \Big) \vec{f}(x,\epsilon),
\end{align}
for the set of ${\cal O}(10^2)$ basis integrals in $\vec f$, where $\alpha_i \in\{x,1+x,1-x,1+x^2\}$ and the $M_i$'s are matrices with rational numbers. The solution can then be cast in terms of MPLs \cite{Duhr:2019tlz}. 
This was the case in all but a $3\times3$ diagonal block, for which it is not possible to achieve the $\epsilon$-factorized form through an algebraic transformation. This turns out to be an elliptic-type sector, which we solved as follows. Firstly, we bring the elliptic diagonal block into the form $A_0 \epsilon^0+A_1 \epsilon^1 $, while transforming all other diagonal blocks into a canonical representation. We~use an ansatz (see e.g.\,\cite{Meyer:2016zeb,Meyer:2016slj}) to make all off-diagonal blocks polynomial~in~$\epsilon$, containing {\it at most} simple poles in $x$, and then we integrate out the $A_0$ component. The last step can be achieved by first deriving the (third-order) Picard-Fuchs equation of the corresponding scalar integral \cite{Primo:2017ipr}.  The~differential equations then contain only simple poles, with the elliptic diagonal block turned into polynomials of degree four~in \beq \{\mathrm{K}(x^2),\mathrm{K}(1-x^2),\mathrm{E}(x^2),\mathrm{E}(1-x^2)\}\,,\label{ell}\eeq where $\mathrm{K},\mathrm{E}$ are the complete elliptic integrals of the first and second kind (our conventions are summarized in the appendix).  
From here it is straightforward to iteratively solve for the master integrals, order-by-order~in~$\epsilon$. This leads to iterated integrals over polylogarithmic kernels as well as elliptic integrals, which can be ultimately re-written in terms of logarithms, dilogarithms and products of $\mathrm{E}$ and $\mathrm{K}$. Finally, we input the boundary (static) values by expanding around $x=1$. It turns out the same seven master integrals used in the (111) sector are sufficient to cover the boundary conditions. Remarkably, similarly to the 3PM case \cite{3pmeft}, this means the test-particle limit carries all the needed information for the boundary terms in the entire potential region. 

\vskip 4pt  Garnering all the pieces together, using \eqref{angle} and \eqref{pmexp}, we arrive at 
\beq
\begin{aligned}
{\chi^{(4)}_{b\, (\rm pot)} \over \pi \Gamma}=  \chi_s(x) + \nu\left( \frac{\chi_{2\epsilon}(x)}{2\epsilon} + \chi_p(x)\right) \,,\label{potangle}
\end{aligned}
\eeq
where $\Gamma \equiv E/M$, with $M$ and $E$ the total mass and energy, and $\nu $ the symmetric mass-ratio, respectively. The result depends on MPLs up to weight two as well as products of the set in \eqref{ell}. The values for the $\left(\chi_s,\chi_{2\epsilon},\chi_p\right)$ coefficients can be found in the appendix. The scheme-dependence appears through the IR pole and renormalization scale, $\bar\mu^2$, as well as various spurious finite pieces which must cancel out altogether against similar ones in the tail contribution \cite{apparent,nrgr4pn2}. Using several identities between MPLs and iterated elliptic integrals, we find our result in~\eqref{potangle} is equivalent to the one obtained in \cite{4pmzvi}, including also the spurious terms. We elaborate on the PM integration and implementation of the differential equations in the EFT approach elsewhere. 
\vskip 4pt
{\it Boundary-to-Bound correspondence.} The scattering angle allows us to derive the bound radial action via the B2B map. Unfortunately, the spurious pieces from the potential region still pollute the answer. However, as a proxy to the full local contribution, we may use the IR-finite part of $\chi^{(\rm 4PM)}_{(\rm pot)}$ to build the (local) bound radial action. Hence, following the B2B correspondence we find for the (IR-finite) correction from potential modes \cite{b2b3}
\begin{align}
 i^{\rm 4PM}_{r \rm (pot)}  =   \frac{2 (1-\gamma^2)^2 }{3(\Gamma j)^3} \left(\chi_s + \nu \chi_p\right) \quad \rm (bound)\,,\label{potir}
\end{align}
where $j \equiv J/(GM^2\nu)$, with $J$ the angular momentum. \vskip 4pt
{\it Impetus \& Hamiltonian.}
The presence of a divergence requires a modification of the Firsov formula discussed~in~\cite{paper1}. Introducing the PM expansion in isotropic gauge for the center-of-mass momentum,
\begin{equation}
  \bp^2(r) = p_\infty^2\left[1+\sum_{n=1}^\infty f_n(E)\left(\frac{G M}{r}\right)^n \left(4 \bar\mu^2 r^2\right)^{n\epsilon}\right]\,,\label{pcom}
\end{equation}
we can then relate the $f_n$'s to the $\chi^{(n)}_b$'s \cite{b2b3}. In particular, at 4PM we have 
\bea
 f_4 &=& \frac{2 (4 \epsilon -1)^3 \Gamma^4 \left(\frac{1}{2}-\epsilon \right)}{3 \pi ^2 \Gamma^4 (1-\epsilon )}\left(\chi_b^{(1)}\right)^4
\label{f4}   \\ &+& \frac{(2-8 \epsilon )^2 \Gamma (1-2 \epsilon ) \Gamma^2 \left(\frac{1}{2}-\epsilon \right)}{\pi ^{3/2} \Gamma \left(\frac{3}{2}-2 \epsilon \right) \Gamma^2 (1-\epsilon )}\left(\chi_b^{(1)}\right)^2 \chi_b^{(2)}\nn \\
    &+&\frac{4 (4 \epsilon -1) \Gamma \left(\frac{3}{2}-3 \epsilon \right) \Gamma   \left(\frac{1}{2}-\epsilon \right)}{\pi  \Gamma (2-3 \epsilon ) \Gamma (1-\epsilon )}\chi_b^{(1)} \chi_b^{(3)}\nn \\
    &+& \frac{ (8 \epsilon -2) \Gamma^2 (1-2 \epsilon )}{\pi  \Gamma^2 \left(\frac{3}{2}-2 \epsilon \right)}\left(\chi_b^{(2)}\right)^2+\frac{2  \Gamma (2-4 \epsilon )}{\sqrt{\pi } \Gamma \left(\frac{5}{2}-4 \epsilon \right)}\chi_b^{(4)}\nn\,,
\eea
where the lower order coefficients can be found in \cite{pmeft,3pmeft}. From here it is straightforward to extract a Hamiltonian. Once again, we write the PM expansion in isotropic gauge as (using a somewhat different convention w.r.t. \cite{4pmzvi})
\beq
H_{\rm (pot)} (\br,\bp^2) = \sum_{n=0}^\infty \frac{c_n(\bp^2)}{n!}\left(\frac{G}{r}\right)^n \left(4\bar\mu^2 r^2\right)^{n\epsilon}\label{4pmH}\,.
\eeq 
The value of $c_4(\bp^2)$ is obtained directly from $f_4(E)$ following the steps described in \cite{paper1}, which remain unaltered, and using
\beq \gamma = \frac{E_1E_2 + \bp^2}{m_1 m_2},\, E_a = \sqrt{m_a^2+\bp^2}\,.\label{rel}\eeq
We find agreement with the result in \cite{4pmzvi}. Complete expressions are given in the appendix.\vskip 4pt  Due to the presence of divergences, the 4PM coefficient of the Hamiltonian acquires the following structure 
\beq
    c_4=-\frac{32M^5(\gamma^2-1)\nu^2}{\Gamma^2\xi}\left(\chi_s+ \nu \chi_p  \right)+ c_4^{(2\epsilon)}+\cdots\,,\label{c4fin}
\eeq
where $\xi = E_1E_2/E^2$, the ellipses account for iterations from lower order terms, and
\beq
c_4^{(2\epsilon)} \equiv -\frac{32M^5(\gamma^2-1)\nu^3}{\Gamma^2\xi}
 \left(\frac{1}{2\epsilon}+
    \frac{10}{3}-4\log(2)\right)\chi_{2\epsilon}\label{c4fin2}\,.
    \eeq
The~contribution due to $c_4^{(2\epsilon)}$ descends directly from the (universal) divergent part of the scattering angle together with the ${\cal O}(\epsilon)$ terms in \eqref{f4}, and therefore is spurious. Hence, it must be removed prior to using potential modes as a proxy of the local Hamiltonian. Unfortunately, other (non-universal) spurious pieces remain in the answer, which are only removed by tail terms \cite{tail4pm,zvitail}.  Luckily, the unambiguous logarithmic part of the full Hamiltonian can be readily computed, as we discuss below. \vskip 4pt 

{\it A prelude for radiation modes.}  The inclusion of radiation modes entails adding {\it self-energy} corrections to the list of Feynman diagrams in Fig.\,\ref{pot}. This corresponds to diagrams with all the sources on the same worldline.\footnote{In principle we must also extend the EFT approach to the (casual) Keldysh-Schwinger formalism \cite{tail}. However, for tail terms it is sufficient to consider the Feynman prescription while keeping the real part of the effective action \cite{hered,foffa3}. This allows us to  retain the integration machinery~intact \cite{tail4pm}.} (Notably only from tree-level and one-loop, with the others yielding scaleless integrals on unperturbed solutions.) The main difference with respect to the potential region is the need to compute integrals involving on-shell graviton modes. Given the nature of the boundary integrals in the soft region, e.g.\,\cite{DiVecchia:2021bdo}, and the expected cancelation of intermediate divergences alongside the factors of $\log \bar \mu^2 b^2$ \cite{tail}, we already anticipate the following structure 
\beq
\begin{aligned}
{\chi^{(4)}_{b\, (\rm rad)} \over \pi \Gamma}=  \nu\left( -\frac{\chi_{2\epsilon}(x)}{2\epsilon} (1-x)^{-4\epsilon} + \chi_t(x)\right) \,,\label{radangle}
\end{aligned}
\eeq 
for the radiation-reaction part of the (conservative) scattering angle.\footnote{The singular factor of $(1-x)^{-4\epsilon}$ signals the existence of (two) quadratic propagators going on-shell in the tail terms. Similarly to $\chi_p(x)$, where intermediate singularities cancel out, $\chi_t(x)$ is likewise devoid of logarithmic divergences in the $x\to 1$ limit~\cite{tail4pm}.} The final result thus takes the form
\beq
\begin{aligned}
\frac{\chi^{(4)}_{b\, (\rm tot)}}{\pi \Gamma} =   \chi_s + \nu \Big(\chi_p +\chi_t +2\chi_{2\epsilon}\log (1-x)\, \Big) \,.\label{totangle}
\end{aligned}
\eeq
The contribution from tail effects to $\chi_t(x)$, which includes physical as well as spurious terms similar to those appearing in $\chi_p(x)$, is reported in \cite{tail4pm}. Nonetheless, we can already derive a series of new results from the structure revealed~in~\eqref{totangle}. \vskip 4pt

{\it Instantaneous flux and total radiated energy.} The coefficient of the pole in the effective action due to tail terms is related to the radiated flux \cite{tail,tail3,Bini:2017wfr,b2b3}. In particular, we have
\beq
\begin{aligned}
S_{\rm eff}^{\rm rad} = - \int_{-\infty}^{+\infty} dt\, H_{\rm( tail)}  =  \frac{GE}{2\epsilon} \int_{-\infty}^{+\infty} dt \frac{dE}{dt} +\cdots\,.
\end{aligned}
\eeq
From the cancelation of divergences, using \eqref{c4fin2} and \eqref{rel}, we can then read off the instantaneous flux in the center-of-mass frame 
\bea
\left. \frac{dE}{dt}\right|_{\rm 3PM}(r,\bp^2)&=& -\frac{4}{3}\frac{G^3 M^4}{r^4}\frac{(\gamma^2-1) \nu^3}{ 
\Gamma^3\xi}\chi_{2\epsilon}(\gamma)\,.\label{dotE} 
\eea
It is straightforward to obtain the total radiated energy, $\Delta E_{\rm hyp/ell}$, both in hyperbolic- and elliptic-like motion, by a change of variables, $dt = dr/\dot r$, and integrating between the end points of the motion.
For instance,
\beq
\Delta E^{\rm 3PM}_{\rm hyp} = -\frac{2\pi \nu^2}{3} \frac{(\gamma^2-1)^2M}{\Gamma^4 j^3}   \chi_{2\epsilon}(\gamma) +\cdots,
\eeq
which agrees (modulo a pre-factor) with the coefficient of the pole in the (unbound) radial action. This reproduces the result obtained in \cite{4pmzvi,Parra2,DiVecchia:2021bdo}. Similarly to the link between angle and periastron advance, the relation between orbital elements discovered in \cite{paper1,paper2}, together with the invariance of the instantaneous flux and radial momentum under $j \to -j$, readily implies 
\beq
\Delta E_{\rm ell}(j) = \Delta E_{\rm hyp}(j)- \Delta E_{\rm hyp}(-j),
\eeq
after analytic continuation  \cite{b2b3} (see also \cite{binidam2}).\vskip 4pt

{\it Logarithmic tail terms.} The universal structure in \eqref{totangle} allows us to compute the logarithmic corrections to the unbound/bound radial action(s). Using  $\log (1-x) \simeq \log v_\infty + \cdots$, with $v_\infty \equiv \sqrt{\gamma^2-1}$, we find 
\beq
\begin{aligned}
{\cal I}^{\rm 4PM}_{r (\rm log)}&= -\frac{E}{(2\pi)M^2\nu}  \Delta E_{\rm hyp}(j) \log v^2_\infty  \quad \rm (unbound)\\
&=   \frac{\nu}{3} \frac{(\gamma^2-1)^2}{(\Gamma j)^3}   \chi_{2\epsilon}(\gamma)  \log v^2_\infty +\cdots \,,
 \end{aligned}
\eeq
defined here so that $\chi/(2\pi) = - \partial_j{\cal I}_r$. We checked that this expression reproduces all the known $ v_\infty^{2n} \log v_\infty^2/j^4$ PN contributions to the non-local part of the scattering angle to 6PN reported in \cite{binidam2}, providing at the same time an infinite series of new terms at ${\cal O}(G^4)$ and to all orders in velocities.\vskip 4pt Remarkably, the same {\it local} B2B dictionary also applies to logarithmic contributions, yielding \cite{b2b3}
\beq
\begin{aligned}
i^{\rm 4PM}_{r (\rm log)}&= -\frac{E}{(2\pi)M^2\nu}  \Delta E_{\rm ell}(j) \log (-{\cal E}) \quad \rm (bound)\\
&=   \frac{2\nu}{3} \frac{(1-\gamma^2)^2}{(\Gamma j)^3}   \chi_{2\epsilon}(\gamma)  \log (-{\cal E}) +\cdots \,,
\end{aligned}
\eeq
with ${\cal E}\equiv (E-M)/(M\nu)<0$, the (reduced) binding energy. The above expression can be then added to \eqref{potir} to describe bound orbits. From here we can obtain, for instance, the periastron advance via $\Delta\Phi/(2\pi) = - \partial_j i_r$. The bound radial action can also be used to compute the functional relationship ${\cal E}(\Omega)$ for circular orbits, with~$\Omega$ the orbital frequency. We have checked that using the condition $i_r=0$ together with Firsov's formula \cite{paper1} reproduces the known PN factors of $\log \Omega$, see~\cite{b2b3}.\vskip 4pt 

{\it Universal part of the non-local Hamiltonian.} The logarithmic contribution to the unbound radial action takes the following form~(see~e.g.~\cite{Bini:2017wfr,nrgr4pn2,binidam2,b2b3}) 
\bea
2\pi GM\, {\cal I}^{\rm 4PM}_{r \rm (nloc)} &=&  - \frac{GE}{M\nu} \bigg(\int_{-\infty}^{+\infty} \frac{d\omega}{2\pi} \frac{dE}{d\omega} \log \left(4\omega^2 e^{2\gamma_E}\right) \nn \\
&&\quad\quad\quad + \int_{-\infty}^{+\infty} dt \frac{dE}{dt} \log r^2(t)\bigg)\nn\\
&=& - \int_{-\infty}^{+\infty} \frac{H_{\rm (nloc)}}{M\nu} dt\,, \label{nonlocal}
\eea
with the factor of $\log 4e^{2\gamma_E}$ introduced following the conventions in the PN literature. From here, using \eqref{dotE} and \eqref{rel}, we derive the logarithmic (non-local) part of the full Hamiltonian at 4PM in isotropic gauge,
\bea
\frac{1}{M\nu} H^{\rm circ}_{\rm (nloc)}(r,\bp^2) &=&   \frac{G\Gamma}{\nu} \frac{dE}{dt}\times \left( \log(\Omega^2 r^2) +\log (16e^{2\gamma_E})\right)\nn\\
&=&  
-\frac{4}{3}\left(\frac{GM}{ r}\right)^4\frac{(\gamma^2-1) \nu^2 }{ \Gamma^2\xi}\chi_{2\epsilon}(\gamma)\nn\\
&&\times \left( \log v^2 +\log (16e^{2\gamma_E})\right)\label{hamlog} \,,
\eea
evaluated on a circular orbit, with $\Omega$ the orbital frequency and $v = \Omega r$. We have added also the constant term on the leading quadrupolar mode $\omega \simeq 2\Omega$. 
\vskip 4pt
{\it Conclusions.} Using the EFT formalism \cite{pmeft,3pmeft} and B2B correspondence \cite{paper1,paper2}, we have computed the contribution from potential modes to the dynamics of non-spinning binary systems to 4PM order, both for hyperbolic- and elliptic-like motion. Our result for the potential-only deflection angle and Hamiltonian agree with the value reported in \cite{4pmzvi} from scattering amplitudes, thus providing an independent derivation.\vskip 4pt 

One of the virtues of the EFT  framework with worldline external sources is that it already systematically encodes all of the relevant (classical) information from the onset. This has the added value of reducing the associated integrals directly into a basis of the type in~\eqref{4pm-ints}, without the need of further manipulations. (See e.g.~\cite{Brandhuber:2021eyq} for an alternative route.) Moreover, as we demonstrated here, the EFT approach is well suited to take full advantage of powerful tools from particle physics to solve the true challenge of a PM scheme---namely the (classical) integration problem---including the method of differential equations \cite{parra,3pmeft}. In particular, the latter gave us full control of the master integrals (to the required order in $\epsilon$), without having to resort to the PN-type resummations in~\cite{zvi1,zvi2,4pmzvi}, which may prove advantageous as we progress towards higher orders.
 \vskip 4pt 

In order to complete the knowledge of the dynamics at 4PM, thus removing spurious terms, self-energy diagrams must be added to the list of topologies as well as regions of integration in the boundary conditions of the differential equations including radiation modes. The full derivation is presented in \cite{tail4pm}, see also \cite{zvitail}. Yet, several contributions can be already obtained from the knowledge of the potential-only result together with the universal character of the non-local tail terms \cite{tail3,b2b3}, as well as the generic properties of the entire soft region with radiation modes, e.g. \cite{DiVecchia:2021bdo}. In particular, we readily derived the logarithmic contributions to the full bound/unbound radial action at~4PM. Through the connection with tail terms, we have also obtained the instantaneous flux at~${\cal O}(G^3)$. The latter allowed us to derive the logarithmic part of the non-local tail Hamiltonian at 4PM. All these results are consistent with the PN values in the literature \cite{binidam2}, while at the same time predict a series of corrections at all orders in velocities. 
In combination with the instantaneous flux in \eqref{dotE}, this provides new ingredients to derive the GW phase evolution entirely within a PM scheme, thus continuing the quest towards high-precision modeling in the era of GW astronomy. \vskip 8pt

{\it Acknowledgements.} We would like to thank Ekta Chaubey, Gihyuk Cho, Stefano Foffa, Ryusuke Jinno, Julio Parra-Martinez, Henrique Rubira and Yang Zhang for useful discussions. 
We also thank the organizers and participants of the workshop ``Gravitational scattering, inspiral, and radiation," at the Galileo Galilei Institute for Theoretical Physics in Florence, for hospitality and fruitful exchanges. G.K., Z.L.~and  R.A.P.~are supported by the ERC-CoG  {\it Precision Gravity: From the LHC to LISA} provided by the European Research Council (ERC) under the European Union's H2020 research and innovation programme (grant No.\,817791).  
The research of C.D. has received support from the ERC under the European Union's Horizon 2020 research and innovation programme, {\it Novel structures in scattering amplitudes} (Grant agreement No.\,725110).

\appendix
\section{Appendix}
Except for the test-body limit, which can be computed from the deflection angle in a Schwarzschild background,
\bea
    \chi_s(x) &=&  \frac{h_1(x)}{128 \left(x^2-1\right)^4}
    =\frac{105 \left(33 \gamma ^4-18 \gamma ^2+1\right)}{128 \left(\gamma ^2-1\right)^2}\label{chis}  \,,
\eea
the remaining terms in the 4PM coefficient of the scattering angle in \eqref{potangle}
can be written in terms of MPLs, $G(\vec{a}; x)$ (up to weight two), times polynomials, $h_i(x)$, in the variable~$x$ as follows:
\begin{widetext}
\begin{equation}
  \begin{aligned}
    \chi_{2\epsilon}(x) = &-\frac{h_2(x)}{64 x^2 \left(x^2-1\right)^4} -\frac{3 h_3(x) G(0;x)}{32 x \left(x^2-1\right)^5}+\frac{3 h_4(x) (G(-1;x)-G(0;2))}{32 x^2 \left(x^2-1\right)^2}\\
    = &-\frac{210 \gamma ^6-552 \gamma ^5+339 \gamma ^4-912 \gamma ^3+3148 \gamma ^2-3336 \gamma +1151}{32 (\gamma^2-1)^2}\\&+\frac{3 \left(35 \gamma ^4+60 \gamma ^3-150 \gamma ^2+76 \gamma -5\right) }{16(\gamma^2 -1)}\log \left(\frac{\gamma+1}{2}\right)
    -\frac{3 \gamma  \left(2 \gamma ^2-3\right) \left(35 \gamma ^4-30 \gamma ^2+11\right) }{32 \left(\gamma ^2-1\right)^{2}}\frac{\arccosh(\gamma )}{\sqrt{\gamma^2-1}}\\
    \chi_p(x) = &-\frac{21 h_5(x) \mathrm{E}^2\left(1-x^2\right)}{8 \left(x^2-1\right)^4}
    +\frac{3 \mathrm{K}\left(1-x^2\right) h_6(x) \mathrm{E}\left(1-x^2\right)}{8\left(x^2-1\right)^4}
    -\frac{15 \mathrm{K}^2\left(1-x^2\right) h_7(x)}{16 \left(x^2-1\right)^4}\\
    &+\frac{\pi ^2 h_8(x)}{256 x^2 \left(x^2-1\right)^5}
    -\frac{h_9(x)}{1536 x^3 \left(x^2-1\right)^6 \left(x^2+1\right)^7}
    +\frac{G(0;2) h_{10}(x)}{128 x^3 \left(x^2-1\right)^4}
    +\frac{3 G(0;2)^2 h_{11}(x)}{16 x^2 \left(x^2-1\right)^2}\\
    &+\frac{h_{12}(x) G(0;x)}{128 x^3 \left(x^2-1\right)^7}
    -\frac{h_{13}(x) G(-1;x)}{128 x^3 \left(x^2-1\right)^4}
    -\frac{h_{14}(x) G\left(-1;x^2\right)}{64 x^3 \left(x^2-1\right)^4}
    -\frac{3 h_{15}(x)G(0;2) G(-1;x)}{64 x^4 \left(x^2-1\right)^2}\\
    &+\frac{3 h_{16}(x) G(0;2)  G(1;x)}{64 x^4\left(x^2-1\right)^2}
    +\frac{3  h_{17}(x) (G(0;x) G(1;x)-G(0,1;x))}{256 x^4 \left(x^2-1\right)^5}\\
    &-\frac{3 h_{18}(x) (G(-1;x) G(1;x)-G(-1,1;x))}{128 x^4 \left(x^2-1\right)^2}
    +\frac{3 h_{19}(x) (G(0;2)  G(0;x)-G(0,-1;x))}{8 \left(x^2-1\right)^5}\\
    &+\frac{3 h_{20}(x) G(0,0;x)}{32 x \left(x^2-1\right)^8}
    -\frac{3 h_{21}(x) G(-1,0;x)}{256 x^4 \left(x^2-1\right)^5}
    +\frac{3 h_{22}(x) G(-1,-1;x)}{128 x^4\left(x^2-1\right)^2}\\
    &-\frac{3 h_{23}(x) \left(G\left(-1;x^2\right) G(1;x)-G(-1,-i;x)-G(-1,i;x)-G(-i,1;x)-G(i,1;x)\right)}{128 x^4 \left(x^2-1\right)^2}\,.  \end{aligned}
  \label{chip}
\end{equation}
We use the following conventions for the complete elliptic integral of the first kind,
\begin{align}\label{}
\mathrm{K}(x^2) = \int_0^{\pi \over 2} \frac{d\theta}{\sqrt{1-x^2 \sin^2\theta}} 
= \int_0^1 \frac{dt}{\sqrt{\left(1-t^2\right)\left(1-x^2 t^2\right)}}
\end{align}
while for the complete elliptic integral of the second kind,
\begin{align}\label{}
\mathrm{E}(x^2) = \int_0^{\pi \over 2} d\theta\, \sqrt{1-x^2 \sin^2\theta} 
= \int_0^1 dt\, \frac{\sqrt{1-x^2 t^2}}{\sqrt{1-t^2}}\,.
\end{align}
The MPLs themselves can be expressed as follows (with $0 < x \leq 1^-$):
\beq
  \begin{aligned}
    G(0;x) = \log (x) \,, \quad\quad\quad\quad\, G(-1;x) &= \log (1+x)\,, \quad\quad\quad\quad G(1;x)= \log(1-x)\,, \\
    G(0,0;x)= \frac{1}{2} \log^2(x)\,, \quad G(-1,-1;x)&=\frac{1}{2} \log^2(1+x)\,,\quad G(0,-1;x)=- \text{Li}_2(-x)\,,\\
    G(-i;x)+G(i;x)= \log(1+x^2)\,, \quad & \quad\quad G(-1,0;x) = \log(x) \log(1+x)+ \text{Li}_2(-x)\,,\\
	G(0,1;x)&=-\frac{\pi^2}{6}+\log(1-x) \log(x)-\frac{\log^2(x)}{2}-\text{Li}_2\left(\frac{x-1}{x}\right)\,,\\
	G(-1,1;x)&=-\frac{\pi^2}{12}+\log(4-4 x) \log(1+x)-\frac{1}{2} \log^2(1+x)\\ &\quad - \log(2) \log(1-x^2) -\text{Li}_2\left(\frac{x-1}{x+1}\right)\,,\\
	G(-1,-i;x)+G(-1,i;x)+G(-i,1;x)+G(i,1;x)&=-\frac{\pi^2}{8}+\log(1+x) \log\left(\frac{4}{1+x^2}\right)+\log(1-x^2) \log\left(\frac{1+x^2}{2}\right)\\ 
	&\quad\,+ \frac{1}{2}  \text{Li}_2\left(-\frac{(x-1)^2}{(1+x)^2}\right)-2 \text{Li}_2\left(\frac{x-1}{x+1}\right)\,. 
  \end{aligned}
\eeq

The $h_i(x)$ polynomials are given by:
\beq
\begin{aligned}
h_1 &= 105 \left(33 x^8+60 x^6+70 x^4+60 x^2+33\right)\\
   h_2 &= 105 x^{12}-552 x^{11}+1308 x^{10}-6408 x^9+29471 x^8-69840 x^7+93368 x^6-69840 x^5+29471
   x^4\\ &\quad -6408 x^3+1308 x^2-552 x+105\,, \\
   h_3 &= 35 x^{13}+60 x^{12}-325 x^{11}+304 x^{10}+198 x^9-788 x^8+446 x^7-889 x^5+788 x^4\\ &\quad -217
   x^3-304 x^2+240 x-60\,,\\
   h_4 &= 35 x^8+120 x^7-460 x^6+968 x^5-1070 x^4+968 x^3-460 x^2+120 x+35\,,\\
   h_5 &= x \left(95 x^4+359 x^2+95\right)\,,\\
      h_6 &= x \left(x^2+1\right) \left(300 x^4+3529 x^2+300\right)\,,\\
      h_7 &= x^3 \left(393 x^4+980 x^2+393\right)\,,\\
      h_8 &= 315 x^{14}-2040 x^{13}+8835 x^{12}-10336 x^{11}-3811 x^{10}+31112 x^9-55627 x^8+33408
   x^7-4099 x^6-3560 x^5\\ &\quad +933 x^4+3040 x^3-3925 x^2+600 x+35\,,\\
   h_9 &=  2700 x^{32}-14178 x^{31}+51621 x^{30}-96582 x^{29}+1268874 x^{28}-5317930 x^{27}+18314727
   x^{26}-43731534 x^{25}\\ & \quad +87689876 x^{24}-143624426 x^{23}+192672621 x^{22}-221881534
   x^{21}+192446998 x^{20}-124929618 x^{19}\\ & \quad+25074951 x^{18}+65377306 x^{17} -91326336
   x^{16}+65377306 x^{15}+25074951 x^{14}-124929618 x^{13}\\ & \quad +192446998 x^{12}-221881534
   x^{11}+192672621 x^{10}-143624426 x^9+87689876 x^8\\ & \quad-43731534 x^7+18314727 x^6-5317930
   x^5+1268874 x^4-96582 x^3+51621 x^2-14178 x+2700\,,\\
   h_{10} &= 300 x^{14}+4523 x^{13}+8948 x^{12}-35378 x^{11}+12700 x^{10}+213317 x^9-814524 x^8\\&\quad+876164
   x^7-814524 x^6+213317 x^5+12700 x^4-35378 x^3+8948 x^2+4523 x+300\,,\\
   h_{11} &= 35 x^8+500 x^6+1362 x^4+500 x^2+35\,,\\
   h_{12} &= 225 x^{20}+4523 x^{19}+856 x^{18}-30593 x^{17}-16165 x^{16}+221833 x^{15}-578424
   x^{14}+54271 x^{13}\\&\quad +1304954 x^{12}-1533425 x^{11}+763800 x^{10}+206421 x^9-603146
   x^8-71285 x^7+355672 x^6\\&\quad-111187 x^5+7925 x^4+18354 x^3-6672 x^2-225\,,\\
   h_{13} &= 150 x^{14}+4523 x^{13}+9018 x^{12}-35378 x^{11}+24206 x^{10}+213317 x^9-672350 x^8+876164
   x^7\\&\quad-672350 x^6+213317 x^5+24206 x^4-35378 x^3+9018 x^2+4523 x+150\,,\\
   h_{14} &= \left(x^2+1\right) \left(75 x^{12}-110 x^{10}-5643 x^8-65444 x^6-5643 x^4-110
   x^2+75\right)\,,\\
   h_{15} &= 25 x^{12}+30 x^{10}+1680 x^9-3801 x^8+13552 x^7-9916 x^6+13552 x^5-3801 x^4+1680 x^3+30
   x^2+25\,,\\
   h_{16} &= 25 x^{12}-250 x^{10}+1680 x^9-7801 x^8+13552 x^7-20812 x^6+13552 x^5-7801 x^4+1680
   x^3-250 x^2+25\,,\\
   h_{17} &= 75 x^{18}-835 x^{16}+4320 x^{15}-19348 x^{14}+21888 x^{13}+6156 x^{12}-62496
   x^{11}+109202 x^{10}-44544 x^9\\&\quad -31930 x^8+25760 x^7-2420 x^6-12160 x^5+12348 x^4-2400
   x^3+555 x^2-75u,,\\
   h_{18} &= 25 x^{12}-530 x^{10}+3360 x^9-13721 x^8+27104 x^7-36572 x^6+27104 x^5-13721 x^4+3360
   x^3-530 x^2+25\,,\\
   h_{19} &= 35 x^{12}+155 x^{10}-26 x^8-1282 x^6+839 x^4+7 x^2-240\,,\\
   h_{20} &= 105 x^{19}-60 x^{18}-90 x^{17}-124 x^{16}-536 x^{15}+1520 x^{14}-1934 x^{13}-3216
   x^{12}+9790 x^{11}+1880 x^{10}\\ & \quad -11350 x^9+1880 x^8+7048 x^7-3216 x^6+94 x^5+1520
   x^4-1559 x^3-124 x^2+480 x-60\,,\\
   h_{21} &= 75 x^{18}+5 x^{16}+4320 x^{15}-13708 x^{14}+21888 x^{13}+5476 x^{12}-62496 x^{11}+71722
   x^{10}-44544 x^9-1538 x^8\\&\quad+25760 x^7-2044 x^6-12160 x^5+5348 x^4-2400 x^3+275 x^2-75\,,\\
   h_{22} &= 25 x^{12}+30 x^{10}+3360 x^9-5721 x^8+27104 x^7-14780 x^6+27104 x^5-5721 x^4+3360 x^3+30
   x^2+25\,,\\
   h_{23} &= 25 x^{12}+30 x^{10}-1881 x^8-5052 x^6-1881 x^4+30 x^2+25\,.
   \end{aligned}
   \eeq

The value of $f_4(E)$ in the expansion of the center-of-mass momentum follows from the scattering angle via 
\bea
 f_4 &=& \frac{2 (4 \epsilon -1)^3 \Gamma^4 \left(\frac{1}{2}-\epsilon \right)}{3 \pi ^2 \Gamma^4 (1-\epsilon )}\left(\chi_b^{(1)}\right)^4
+ \frac{(2-8 \epsilon )^2 \Gamma (1-2 \epsilon ) \Gamma^2 \left(\frac{1}{2}-\epsilon \right)}{\pi ^{3/2} \Gamma \left(\frac{3}{2}-2 \epsilon \right) \Gamma^2 (1-\epsilon )}\left(\chi_b^{(1)}\right)^2 \chi_b^{(2)} \\
    &+&\frac{4 (4 \epsilon -1) \Gamma \left(\frac{3}{2}-3 \epsilon \right) \Gamma   \left(\frac{1}{2}-\epsilon \right)}{\pi  \Gamma (2-3 \epsilon ) \Gamma (1-\epsilon )}\chi_b^{(1)} \chi_b^{(3)}
    + \frac{ (8 \epsilon -2) \Gamma^2 (1-2 \epsilon )}{\pi  \Gamma^2 \left(\frac{3}{2}-2 \epsilon \right)}\left(\chi_b^{(2)}\right)^2+\frac{2  \Gamma (2-4 \epsilon )}{\sqrt{\pi } \Gamma \left(\frac{5}{2}-4 \epsilon \right)}\chi_b^{(4)}\nn\,.
\eea
The 4PM coefficient of the Hamiltonian in isotropic gauge is given by
\begin{equation}
  \begin{aligned}\label{c4}
    c_4(\bp^2)=&\frac{3 \left(\gamma ^2-1\right) M^5 \nu ^8 d_1(\gamma ) f_1(E){}^4}{16 \Gamma ^{27} \xi^7}+\frac{9 \left(\gamma ^2-1\right)^3 M^7 \nu ^8 d_2(\gamma ) f_1(E){}^2 f_1'(E){}^2}{2 \Gamma ^{17} \xi^5}\\
    &-\frac{9 \left(\gamma ^2-1\right)^2 M^6 \nu ^6 d_2(\gamma ) f_1(E) f_2(E) f_1'(E)}{\Gamma ^{14} \xi^4}
    -\frac{9 \left(\gamma ^2-1\right)^2 M^6 \nu ^6 d_2(\gamma ) f_1(E){}^2 f_2'(E)}{2 \Gamma ^{14} \xi^4}\\
    &+\frac{3 \left(\gamma ^2-1\right)^3 M^7 \nu ^8 d_2(\gamma ) f_1(E){}^3 f_1''(E)}{2 \Gamma ^{17} \xi^5}
    +\frac{3 \left(\gamma ^2-1\right)^2 M^6 \nu ^8 d_3(\gamma ) f_1(E){}^3 f_1'(E)}{4 \Gamma ^{22} \xi^6}
    \\
    &+\frac{9 \left(\gamma ^2-1\right) M^5 \nu ^6 d_4(\gamma ) f_1(E){}^2 f_2(E)}{2 \Gamma ^{19} \xi^5}
    +\frac{3 \left(\gamma ^2-1\right) M^5 \nu ^4 d_2(\gamma ) f_2(E){}^2}{\Gamma ^{11} \xi ^3}\\
    &+\frac{6\left(\gamma ^2-1\right) M^5 \nu ^4 d_2(\gamma ) f_1(E) f_3(E)}{\Gamma ^{11} \xi ^3}
    +\frac{9\left(\gamma ^2-1\right)^4 M^8 \nu ^8 f_1(E){}^2 f_1'(E) f_1''(E)}{4 \Gamma ^{12} \xi ^4}\\
    &+\frac{3\left(\gamma ^2-1\right)^4 M^8 \nu ^8 f_1(E) f_1'(E){}^3}{2 \Gamma ^{12} \xi ^4}
    -\frac{3 \left(\gamma^2-1\right)^3 M^7 \nu ^6 f_2(E) f_1'(E){}^2}{\Gamma ^9 \xi ^3}\\
    &-\frac{6 \left(\gamma ^2-1\right)^3 M^7\nu ^6 f_1(E) f_1'(E) f_2'(E)}{\Gamma ^9 \xi ^3}
    +\frac{6 \left(\gamma ^2-1\right)^2 M^6 \nu ^4 f_3(E)f_1'(E)}{\Gamma ^6 \xi ^2}\\
    &+\frac{6 \left(\gamma ^2-1\right)^2 M^6 \nu ^4 f_2(E) f_2'(E)}{\Gamma ^6 \xi^2}
    +\frac{6 \left(\gamma ^2-1\right)^2 M^6 \nu ^4 f_1(E) f_3'(E)}{\Gamma ^6 \xi ^2}\\
    &-\frac{3\left(\gamma ^2-1\right)^3 M^7 \nu ^6 f_1(E) f_2(E) f_1''(E)}{\Gamma ^9 \xi ^3}
    -\frac{3 \left(\gamma^2-1\right)^3 M^7 \nu ^6 f_1(E){}^2 f_2''(E)}{2 \Gamma ^9 \xi ^3}\\
    &+\frac{\left(\gamma ^2-1\right)^4 M^8\nu ^8 f_1(E){}^3 f_1'''(E)}{4 \Gamma ^{12} \xi ^4}
    -\frac{12 \left(\gamma ^2-1\right) M^5 \nu ^2 f_4(E)}{\Gamma ^3 \xi }\,,
  \end{aligned}
\end{equation}
where primes denote derivatives w.r.t. to $E$,  and the $d_i$'s are polynomials in $\gamma$ given by
\begin{equation}
  \begin{aligned}\label{d4}
    d_1(\gamma) = &\left(35 \gamma ^3-135 \gamma ^2+201 \gamma -141\right) \nu ^6 (\gamma -1)^9\\
    &+3 \left(75 \gamma ^4-279 \gamma ^3+377 \gamma ^2-93 \gamma +200\right) \nu ^5 (\gamma -1)^7\\
    &+\left(603 \gamma^5-2115 \gamma ^4+2063 \gamma ^3-1491 \gamma ^2+474 \gamma -934\right) \nu ^4 (\gamma -1)^5\\
    &+\left(855 \gamma ^6-2563 \gamma ^5+2617 \gamma ^4-1749 \gamma ^3+1076 \gamma ^2-668 \gamma +712\right) \nu ^3 (\gamma -1)^3\\
    &+\left(641\gamma ^6-1108 \gamma ^5+811 \gamma ^4-420 \gamma ^3+419 \gamma ^2-152 \gamma +289\right) \nu ^2 (\gamma -1)^2\\
    &+\left(239 \gamma ^6-181 \gamma ^5+134 \gamma ^4-46 \gamma ^3+95 \gamma ^2-13 \gamma +60\right) \nu  (\gamma -1)+35\gamma ^6+15 \gamma ^4+9 \gamma ^2+5\,,\\
    d_2(\gamma) = &\left(7 \gamma ^2-5 \gamma +4\right) (\gamma -1) \nu +3 \gamma ^2+(3 \gamma -5) (\gamma -1)^3 \nu ^2+1\,,\\
    d_3(\gamma) = &\left(29 \gamma ^2-86 \gamma +77\right) \nu ^4 (\gamma -1)^6
    +2 \left(65 \gamma ^3-162 \gamma ^2+81 \gamma-76\right) \nu ^3 (\gamma -1)^4\\
    &+\left(211 \gamma ^4-382 \gamma ^3+305 \gamma ^2-136 \gamma +118\right) \nu ^2 (\gamma -1)^2\\
    &+2 \left(67 \gamma ^4-49 \gamma ^3+45 \gamma ^2-11 \gamma +20\right) \nu  (\gamma -1)
    +29 \gamma ^4+14\gamma ^2+5\,,\\
    d_4(\gamma) = &-\left(5 \gamma ^2-14 \gamma +13\right) \nu ^4 (\gamma -1)^6
    -2 \left(11 \gamma ^3-28 \gamma ^2+11 \gamma -14\right) \nu ^3 (\gamma -1)^4\\
    &-\left(36 \gamma ^4-66 \gamma ^3+47 \gamma ^2-20\gamma +23\right) \nu ^2 (\gamma -1)^2\\
    &-\left(23 \gamma ^4-17 \gamma ^3+13 \gamma ^2-3 \gamma +8\right) \nu  (\gamma -1)
    -5 \gamma ^4-2 \gamma ^2-1\,.
  \end{aligned}
\end{equation}
The final expression for $c_4$ becomes a function of the center-of-mass momentum by replacing
\beq \gamma = \frac{E_1E_2 + \bp^2}{m_1 m_2},\, E_a = \sqrt{m_a^2+\bp^2}\,,
\eeq
after taking the derivatives. The result for potential-only modes coincides with the value reported in \cite{4pmzvi}. 
 \end{widetext}
\bibliography{ref4PM}

\end{document}